\documentclass[a4paper,11pt]{article}
\pdfoutput=1 

\usepackage{jcappub} 

\usepackage[T1]{fontenc} 

\title{$\sigma$CDM coupled to radiation. Dark energy and Universe acceleration}

\author[a]{R.\,R.\,Abbyazov,}
\author[a,b,c]{S.\,V.\,Chervon,}
\author[d]{V.\,M\"uller}

\affiliation[a]{Ilya Ulyanov State Pedagogical University,\\Ulyanovsk 432700, Russia}
\affiliation[b]{Immanuel Kant Baltic Federal University,\\Kaliningrad 236041, Russia}
\affiliation[c]{University of KwaZulu-Natal,\\Durban 4000, South Africa}
\affiliation[d]{Leibniz-Institut f\"ur Astrophysik Potsdam,\\
An der Sternwarte 16, 14482 Potsdam, Germany}

\emailAdd{renren2007@yandex.ru}
\emailAdd{chervon.sergey@gmail.com}
\emailAdd{vmueller@aip.de}

\abstract{Recently the Chiral Cosmological Model (CCM) coupled to cold
dark matter (CDM) has been investigated as $\sigma$CDM model to study
the observed accelerated expansion of the Universe. Dark sector fields
(as Dark Energy content) coupled to cosmic dust were considered as the
source of Einstein gravity in Friedmann-Robertson-Walker (FRW)
cosmology. Such model had a beginning at the matter-dominated era. The
purposes of our present investigation are two folds: to extend
<<life>> of the $\sigma$CDM for earlier times to radiation-dominated
era and to take into account variation of the exponential potential
$V=V_0\exp \left( {-\sqrt{\lambda}\frac{\ph}{M_{P\ell}}}
\right)+V_0\exp \left( {-\sqrt{\lambda}\frac{\chi}{M_{P\ell}}}
\right)$ via variation of the interaction parameter $\lambda $.

We use Markov Chain Monte Carlo (MCMC) procedure to investigate
possible values of initial conditions constrained by the measured
amount of the dark matter, dark energy and radiation component today.

Our analysis includes dark energy contribution to critical density,
the ratio of the kinetic and potential energies, deceleration
parameter, effective equation of state and evolution of DE equation of
state with variation of coupling constant $\lambda $.  A comparison
with the $\Lambda$CDM model was performed.  A new feature of the model
is the existence of some values of potential coupling constant,
leading to a $\sigma$CDM solution without transit into accelerated
expansion epoch.
}

\newcommand{\ph}{\varphi}
\newcommand{\prt}{\partial}

\begin{document}
\maketitle
\flushbottom

\section{Introduction}
\label{sec:intro}

To explain the present accelerated expansion of the universe, confirmed by
a range of astrophysical observations, one may use dark energy (in a
wide sense as the source of the acceleration with negative pressure)
or one has to modify Einstein's gravitation theory. 

The simplest and very attractive model of dark energy is one based on
the cosmological constant $\Lambda$ as a repulsive force
\cite{Tsujikawa:2013fta, Tsujikawa:2010sc, cosats06, Li:2011sd,
Padmanabhan:2002ji}. The resulting $\Lambda$CDM model with cold dark
matter component is consistent with astrophysical observations in many
respects. However, the model is faced with two problems: fine tuning
and cosmic coincidence.

The cosmological constant problem essentially has to do with our
(mis)understanding of the nature of gravity \cite{Padmanabhan:2007xy}.
Other option in understanding of cosmic acceleration and dark energy
is the infrared modification of general relativity that may be
responsible for the large-scale behavior of the universe
\cite{Silvestri:2009hh}.

Also today a very popular approach concerns modified gravity, where
additional degrees of freedom must screen themselves from local tests
of gravity.  So there exist screening mechanisms associated with
chameleon and galileon mechanisms, as well as massive gravity and
Vainshtein mechanisms (see, for details \cite{Joyce:2014kja}). Various
observational and experimental tests have been reflected in this work
as well.

Another approach to avoid fine tuning and cosmic coincidence problems
is connected with a variety of scalar field models, including
quintessence, phantom's and tachyon's fields etc. have been
proposed. One of such model, so called $\phi$CDM model, is presented
in the work Park et.\,al.  \cite{Hwang:2009zj}.
The k--essence model providing a dynamical solution for explaining
naturally why the universe has entered an epoch of accelerated
expansion at a late stage of its evolution also essentially relies on
using scalar fields. The solution that proposed in k--essence model
avoids fine-tuning of parameters and anthropic arguments 
\cite{ArmendarizPicon:2000ah}.

In the present article we extend in some sense such types of models
to a multiplet of the scalar fields and work with the
two-component chiral cosmological model in details.
 
A chiral cosmological model (CCM) have been proposed as a non--linear
sigma model with the potential of (self)interaction and implemented
into cosmology (for review, see \cite{chervon2013chiral}).
Application of a CCM coupled to a cold dark matter for later time
inflation leads to proposal of a new type of models with quintessence and
quintom fields, dubbed as $\sigma$CDM models
\cite{abbyazov_chervon_12:_inter_chiral_field_dark_sector}.  Also the
cosmological constant problem is absent in CCM as there exist a
dynamic potential interaction in CCM. In this work the necessity of extension 
of this model to more general CCM was mentioned to obtain better
agreement with observational data. On this way we investigate
now an extension of $\sigma$CDM models presented in
\cite{abbyazov_chervon_12:_inter_chiral_field_dark_sector}.  Namely we
start from $\sigma$CDM model that more drastically differs from
$\Lambda$CDM. To go deeper back in time we include radiation in the
model action. Thus our consideration may start from radiation
dominated era at times which roughly corresponds to nucleosynthesis
stage. In our approach we present dark energy as kinetically and
potentially interacting two scalar fields described by a chiral
cosmological model. We also include a coupling to
perfect fluids, cold dark matter and radiation.

Recently authors of \cite{Chiba:2014sda} investigated multiple scalar
fields model, including a model of two canonical fields with coupled
exponential potentials arising in string theory. The model they
investigated is similar to our one, but devoted to cosmological
scaling solutions.

It is also very common to study dark energy models by means of some
parametrization of the dark energy equation of state.  One often used
dark energy parametrization of dark energy is CPL parametrization with
dependence on scale factor $a(t)$ in the form $w(a)=w_0+w_a (1-a)$
\cite{Hazra:2013dsx} (called CPL after Chevallier -- Polarksi --
Linder).  The authors \cite{Hazra:2013dsx} mentioned that almost all
the cosmological constraints on dark energy are based on this
parametrization.  The question still stands whether there are possible
dark energy evolutions that one misses using the CPL parametrization.
Their results motivate the construction of models of dark energy which
lead to phantom behavior. This means going beyond standard
possibilities for dark energy involving a scalar field with a positive
kinetic energy term only which do not lead to a violation of the weak
energy condition \cite{Hazra:2013dsx}.
 
There exist also the Scherrer and Sen parametrization\cite{Sen:2005sk} 
represented by a rather complicated formula
with dependence $w(a)=(1+w_0) f(\Omega_{DE},a)$ with a certain function $f$ 
and the 
Generalized Chaplygin Gas (GCG) parametrization presented by the relation
$
w(a)=-\frac{A}{A+(1-A)a^{-3(1+\alpha)}},~~0 \geqslant \alpha \geqslant 1.
$ 
Thus we can see that equation of state (EoS) may not be constant and could 
vary with the scale factor $a$.

In our presentation the EoS parameter is obtained numerically, and the dependence
on $a$ is displayed graphically.

Authors \cite{Xia:2013dea} attracted attention to the tension between measured the Hubble constant
$H_0$ by the Planck collaboration and by the several direct probes on $H_0$. To
avoid this tension they found out that EoS $w$ should be less the $-1 $ or it
should be time-evolving. Also they mentioned that with such tension
the concordance cosmological model ($\Lambda$CDM) is in fact incomplete. This is
also one more reason to analyze the $\sigma$CDM model.

The general plan of the paper is the following. In the
sec.\,\ref{sec:sec1} we represent the main equations of the 
$\sigma$CDM model under investigation and write down basic
cosmological quantities that can be extracted from that
equations. Also we make a choice of new variables, more
appropriate in the approach connected with usage of initial values of
quantities.  In sec.\,\ref{sec:radiation_inclusion} we describe
observational constraints on $\sigma$CDM which we use here and the MCMC
procedure. In sec.\,\ref{sec:disc} we discuss the dependence of
cosmological dynamics on varied potential interaction parameters
values $\lambda$, make comparison with $\Lambda$CDM model, as well as
point out essential features of $\sigma$CDM model, such as magnitude
of kinetic interaction with relation to potential
one. Sec.\,\ref{sec:conclusion} is devoted for possible extension of
the work we have done in this paper.

\section{Extension of $\sigma$CDM model to radiation dominated era}
\label{sec:sec1}

We start with consideration of a Chiral Cosmological Model (CCM)
coupled to perfect fluid.
Such a model related to (dark) matter source was called $\sigma\mathrm{CDM}$
model
\cite{Abbyazov:2013qqa,abbyazov_chervon_12:_inter_chiral_field_dark_sector}.

Our intention now is to extend the model deeper back on time to the end of
BB nucleosynthesis and beginning of the radiation era.

The action of CCM coupled to perfect fluid is
\begin{equation*}
S=\int d^4x \sqrt{-g}\left( -\frac{1}{2}g^{\mu\nu}h_{AB}\prt_\mu\ph^A\prt_\nu\ph^B-V(\ph^A)  \right)+S_{pf},
\end{equation*}
where $S_{pf}$ accounts for perfect fluid (cold dark matter).
We choose chiral metric components as $h_{11}=1,~h_{22}=\exp\left({\sqrt{\mu} \frac{\ph-\ph_i}{M_{P\ell}}
}\right)$ and the (self)action potential in the form $V=V_0\exp \left(
{-\sqrt{\lambda}\frac{\ph}{M_{P\ell}}} \right)+V_0\exp \left(
{-\sqrt{\lambda}\frac{\chi}{M_{P\ell}}} \right)$, where
$M_{P\ell}\equiv1/\sqrt{8\pi G}$. Also we have to include in the model
(dark and baryon) matter and radiation with dynamics represented by equations
of energy--momentum conservation for each component.

To describe the dynamics of the model we study the system of Friedmann and
chiral cosmological fields equations:

\begin{eqnarray}
  \label{eq:ein_field_eqs_original_friedmann_eq}
  H^2=\frac{8\pi G}{3}\left[ \rho_m+\rho_r+\frac{1}{2}\dot\ph^2
+\frac{1}{2}h_{22}\dot{\chi}^2+V(\ph, \chi)\right],\\
\label{eq:ein_field_eqs_original_1_field_eq}
\ddot{\ph}+3H\dot{\ph}-\frac{1}{2}\frac{\prt h_{22}}{\prt\ph}\dot{\chi}^2+
\frac{\prt V}{\prt\ph}=0,\\
\label{eq:ein_field_eqs_original_2_field_eq}
\ddot{\chi}+3H\dot{\chi}+\frac{1}{h_{22}}\frac{\prt h_{22}}{\prt\ph}\dot{\ph}
\dot{\chi}+\frac{1}{h_{22}}\frac{\prt V}{\prt\chi}=0.
\end{eqnarray}

Let us mention here that the second Einstein's equation (Raichaudhury
equation) can also be derived by using linear combination of chiral
cosmological fields, conservation of matter and radiation, and the
Friedmann equation.

The key idea \cite{Seshadri2008}, that allows us to
solve this ODE system, is to rely on initial values
of quantities presented there. It means, that
energy--momentum tensor conservation gives us
energy densities of matter and radiation components
in the following form

\begin{equation*}
  \rho_r=\rho^{(i)}_{r}\left(\frac{a}{a_i}\right)^{-4}=
\frac{3H^2_i}{8\pi G}\Omega^{(i)}_{r}\left(\frac{a}{a_i}\right)^{-4},
\end{equation*}

\begin{equation*}
  \rho_m=\rho^{(i)}_{m}\left(\frac{a}{a_i}\right)^{-3}=
\frac{3H^2_i}{8\pi G}\Omega^{(i)}_{m}\left(\frac{a}{a_i}\right)^{-3},
\end{equation*}

\begin{equation*}
  \rho^{(i)}_{c}=\frac{3H^2_i}{8\pi G},
\end{equation*}
where we introduced the critical density of the Universe
for some initial moment of time which corresponds to
scale factor value $a=a_i$. This approach
allows us besides to get a feeling about
values of initial amounts of dark energy, matter and
radiation. We can estimate these values from the $\Lambda$CDM
model if we put in corresponding expressions for the 
amounts of dark energy, matter and
radiation today. The latter values should
be in concordance with modern observational data,
therefore we use values $\Omega^{(0)}_{DE}\approx0.7$, $\Omega^{(0)}_{m}\approx0.3$,
$\Omega^{(0)}_r=5\cdot10^{-5}$.

There is a restriction for maximum possible amount of dark energy (or
in general scalar field model contribution) for nucleosynthesis epoch
$\Omega_{DE}<0.045$ \cite{Hwang:2009zj,Bassett:2007aw,Bean:2001wt}.
We take this value as upper bound for dark energy contribution at
initial time.  In our analysis we want to be rather conservative and
let lower bound of the dark energy amount be equal to
$1\cdot10^{-25}$. These values are used for boundary values of $\Omega_{DE}^{(i)}$ 
in MCMC procedure.

According to our choice above we would like to use new type
of dimensionless variables \cite{Seshadri2008}
\begin{equation*}
    s=\frac{a}{a_i},~y=\frac{\ph-\ph_{i}}{M_P},~z=\frac{\chi-\chi_{i}}{M_P}.
\end{equation*}

\begin{equation*}
  x=H_i\left(t-t_i\right), \text{<<'>>}=\frac{d}{dx}, M_P^2=\frac{1}{8\pi G},
\end{equation*}

\begin{equation*}
    \bar V=V_0M_P^{-2}H_i^{-2}\exp\left(-\sqrt{\lambda}M_P^{-1}\ph_{i}\right)=
V_0M_P^{-2}H_i^{-2}\exp\left(-\sqrt{\lambda}M_P^{-1}\chi_{i}\right).
\end{equation*}

Thus the original system of equations is transformed to

\begin{eqnarray}
  \label{eq:ein_field_eqs_original_friedmann_eq}
\left(\frac{s'}{s}\right)^2=\Omega^{(i)}_{m}s^{-3}+\Omega^{(i)}_{m}s^{-4}+
\frac{1}{3}\left[
  \frac{1}{2}y'^2+\frac{1}{2}h_{22}z'^2+\bar{V}e^{-\sqrt{\lambda}y}+\bar{V}e^{-\sqrt{\lambda}z}\right],\\
\label{eq:ein_field_eqs_original_1_field_eq}
y''+3Hy'-\frac{1}{2}\sqrt{\mu}e^{\sqrt{\mu}y}z'^2-\sqrt{\lambda}\bar{V}e^{-\sqrt{\lambda}y}=0,\\
\label{eq:ein_field_eqs_original_2_field_eq}
z''+3Hz'+\sqrt{\mu}e^{\sqrt{\mu}y}y'z'
-\sqrt{\lambda}e^{\sqrt{\mu}y}\bar{V}e^{-\sqrt{\lambda}z}=0.
\end{eqnarray}
The chiral cosmological fields act as dark energy and contributions
to the critical density of the universe and the dark energy equation
of state parameter as

\begin{equation}
  \label{eq:Omega_DE_CCM_new_variables}
  \Omega_{DE}=\frac{\frac{1}{3}\left[
\frac{1}{2}y'^2+\frac{1}{2}h_{22}z'^2+\bar{V}e^{-\sqrt{\lambda}y}+\bar{V}e^{-\sqrt{\lambda}z}
\right]}{\Omega^{(i)}_{m}s^{-3}+
\Omega^{(i)}_{r}s^{-4}+\frac{1}{3}\left[
\frac{1}{2}y'^2+\frac{1}{2}h_{22}z'^2+\bar{V}e^{-\sqrt{\lambda}y}+\bar{V}e^{-\sqrt{\lambda}z}
\right]},
\end{equation}

\begin{equation}
  \label{eq:DE_eq_of_state_param_new_variables}
  \omega_{DE}=\frac{
\frac{1}{2}y'^2+\frac{1}{2}h_{22}z'^2-\bar{V}e^{-\sqrt{\lambda}y}-\bar{V}e^{-\sqrt{\lambda}z}
}{\frac{1}{2}y'^2+\frac{1}{2}h_{22}z'^2+\bar{V}e^{-\sqrt{\lambda}y}+\bar{V}e^{-\sqrt{\lambda}z}}.
\end{equation}

Initial conditions for chiral cosmological fields
are
\begin{equation*}
  y'_{i}=z_{i}'=\sqrt{\frac{3}{2}\left(
      1-\Omega^{(i)}_{m}-\Omega^{(i)}_r \right)
\left( 1-\omega^{(i)}_{DE} \right)},
\end{equation*}
 and for the potential is
\begin{equation*}
\bar{V}=\frac{3}{4}\left( 1-\Omega^{(i)}_{m}-\Omega^{(i)}_r \right)
\left(1-\omega^{(i)}_{DE}\right),
\end{equation*}
It needs to keep in mind that we consider initial contributions
of the both chiral cosmological fields to be equal
$\Omega^{(i)}_{\ph}=\Omega^{(i)}_{\chi}$.

In the present paper we analyze the evolution of such quantities as
deceleration parameter, effective equation of state
parameter and chiral cosmological fields kinetic and potential energies ratio.
We also make a comparison with the corresponding quantities of
the $\Lambda$CDM model. Let us write down expressions for them
in terms of set of variables we introduced previously

\begin{multline}
  \label{eq:q_deceleration_parameter}
  q^{\sigma\text{CDM}}=-\frac{\frac{\ddot{a}}{a}}{\left( \frac{\dot{a}}{a} \right)^2}=
\frac{\frac{4\pi G}{3}\sum_{\alpha}\left( \rho_{\alpha}+3p_{\alpha} \right)}{\frac{8\pi G}{3}\sum_{\alpha}\rho_{\alpha}}=
\frac{1}{2}\frac{
\left[ \rho_{r}+\rho_{m}+\rho_{\sigma}+3p_{r}+3p_{m}+3p_{\sigma} \right]}
{\left[ \rho_{r}+\rho_{m}+\rho_{\sigma} \right]}=\\
=\frac{1}{2}\frac{\left[ 2\Omega^{(i)}_{r}s^{-4}+
\Omega^{(i)}_{m}s^{-3}+\frac{1}{3}\left[ 2y'^2+2h_{22}z'^2-2\bar{V}e^{-\sqrt{\lambda}y}-2\bar{V}e^{-\sqrt{\lambda}z} \right]\right]}
{\left[ \Omega^{(i)}_{r}s^{-4}+\Omega^{(i)}_{m}s^{-3}+
\frac{1}{3}\left[ \frac{1}{2}y'^2+\frac{1}{2}h_{22}z'^2+ \bar{V}e^{-\sqrt{\lambda}y}+ \bar{V}e^{-\sqrt{\lambda}z} \right] \right]},
\end{multline}

\begin{multline}
  \label{eq:omega_eff}
  \omega^{\sigma\text{CDM}}_{eff}=\frac{\sum_{\alpha}p_{\alpha}}{\sum_{\alpha}\rho_{\alpha}}=
\frac{p_{r}+p_{m}+p_{\sigma}}{\rho_{r}+\rho_{m}+\rho_{\sigma}}=\\
=\frac{\frac{1}{3}\Omega^{(i)}_{r}s^{-4}+\frac{1}{3}\left[\frac{1}{2}y'^{2}+\frac{1}{2}h_{22}z'^2
-\bar{V}e^{ -\sqrt{\lambda}y} - \bar{V}e^{ -\sqrt{\lambda}z}\right]}
{\Omega^{(i)}_{r}s^{-4}+\Omega^{(i)}_{m}s^{-3}+\frac{1}{3}\left[ \frac{1}{2}y'^2+\frac{1}{2}h_{22}z'^2
+\bar{V}e^{ -\sqrt{\lambda}y} + \bar{V}e^{ -\sqrt{\lambda}z }\right]},
\end{multline}

\begin{equation}
  \label{eq:K_over_V_ratio}
  \frac{K}{V}=\frac{\frac{1}{2}y'^{2}+\frac{1}{2}h_{22}z'^2}
{\bar{V}e^{ -\sqrt{\lambda}y} + \bar{V}e^{ -\sqrt{\lambda}z }}.
\end{equation}

For the $\Lambda$CDM model we have

\begin{multline}
  \label{eq:q_deceleration_parameter_LCDM}
  q^{\Lambda\text{CDM}}=-\frac{\frac{\ddot{a}}{a}}{\left( \frac{\dot{a}}{a} \right)^2}=
\frac{\frac{4\pi G}{3}\sum_{\alpha}\left( \rho_{\alpha}+3p_{\alpha} \right)}{8\pi G/3\sum_{\alpha}\rho_{\alpha}}=\\
=\frac{1}{2}\frac{
\left[ \rho_{r}+\rho_{m}+\rho_{\Lambda}+3p_{r}+3p_{m}+3p_{\Lambda} \right]}
{\left[ \rho_{r}+\rho_{m}+\rho_{\Lambda} \right]}=
\frac{1}{2}\frac{\left[ 2\Omega^{(i)}_{r}s^{-4}+
\Omega^{(i)}_{m}s^{-3}-2\Omega^{(i)}_{\Lambda}\right]}
{\Omega^{(i)}_{r}s^{-4}+\Omega^{(i)}_{m}s^{-3}+
+\Omega^{(i)}_{\Lambda}},
\end{multline}

\begin{equation}
  \label{eq:omega_eff_LCDM}
  \omega^{\Lambda\text{CDM}}_{eff}=\frac{\sum_{\alpha}p_{\alpha}}{\sum_{\alpha}\rho_{\alpha}}=
\frac{p_{r}+p_{m}+p_{\Lambda}}{\rho_{r}+\rho_{m}+\rho_{\Lambda}}
=\frac{\frac{1}{3}\Omega^{(i)}_{r}s^{-4}-\Omega^{(i)}_{\Lambda}}
{\Omega^{(i)}_{r}s^{-4}+\Omega^{(i)}_{m}s^{-3}+\Omega^{(i)}_{\Lambda}}.
\end{equation}

\section{Constraining and solving $\sigma$CDM equations}
\label{sec:radiation_inclusion}

We have solved background  Einstein and scalar field equations of
$\sigma\mathrm{CDM}$ model for $\lambda$ values lying between 0.1 and
10 while keeping kinetic interaction coupling constant fixed and equal
to $\mu=1.0$. $\lambda$ values are choosen as
\[\lambda=\left\{0.10, 0.17, 0.28, 0.46, 0.77, 1.29, 2.15, 3.59, 5.99, 10.0  \right\}.\]
Initial contribution of matter and dark energy (chiral cosmological fields) in early epoch $a_i=10^{-6}$
tuned according to MCMC algorithm so that one can get current amount
of dark energy today equal to 0.7
and current amount of radiation today equal to $5\cdot10^{-5}$.

From the earlier work
\cite{abbyazov_chervon_12:_inter_chiral_field_dark_sector} one can see
that there are a substantial deviation of dark energy equation of
state parameter $\omega_{DE}$ from $-1$ value even in model without
radiation and rather late times. Our next step in this direction is
finding possible nonnegligible dark energy contribution in early
epochs as it take place in tracker models. So we try to take into
consideration a radiation component in order to follow the more early
universe evolution.

In order to fit our model to present day observations we propose target
function to be minimised by MCMC procedure. We take it to be

\begin{equation*}
  \chi^2_{joint}=\sqrt{\chi^2_{DE}\chi^2_{DE}+\chi^2_{r}\chi^2_{r}},
\end{equation*}
where
\begin{equation*}
  \chi^2_{DE}=\left(\Omega^{(0)}_{DE}-\Omega^{b.f.}_{DE}\right)/\Omega^{b.f.}_{DE},
\end{equation*}

\begin{equation*}
  \chi^2_{r}=\left(\Omega^{(0)}_{r}-\Omega^{b.f.}_{r}\right)/\Omega^{b.f.}_{r},
\end{equation*}
and $\Omega^{b.f.}_{DE}=0.7$, $\Omega^{b.f.}_{r}=5\cdot 10^{-5}$.

As initial contributions to critical density varies very
broadly in orders we consider their decadic logarithm values.

\begin{figure}[h!]
\begin{minipage}[v]{0.5\linewidth}
\center{\includegraphics[width=1.0\linewidth]{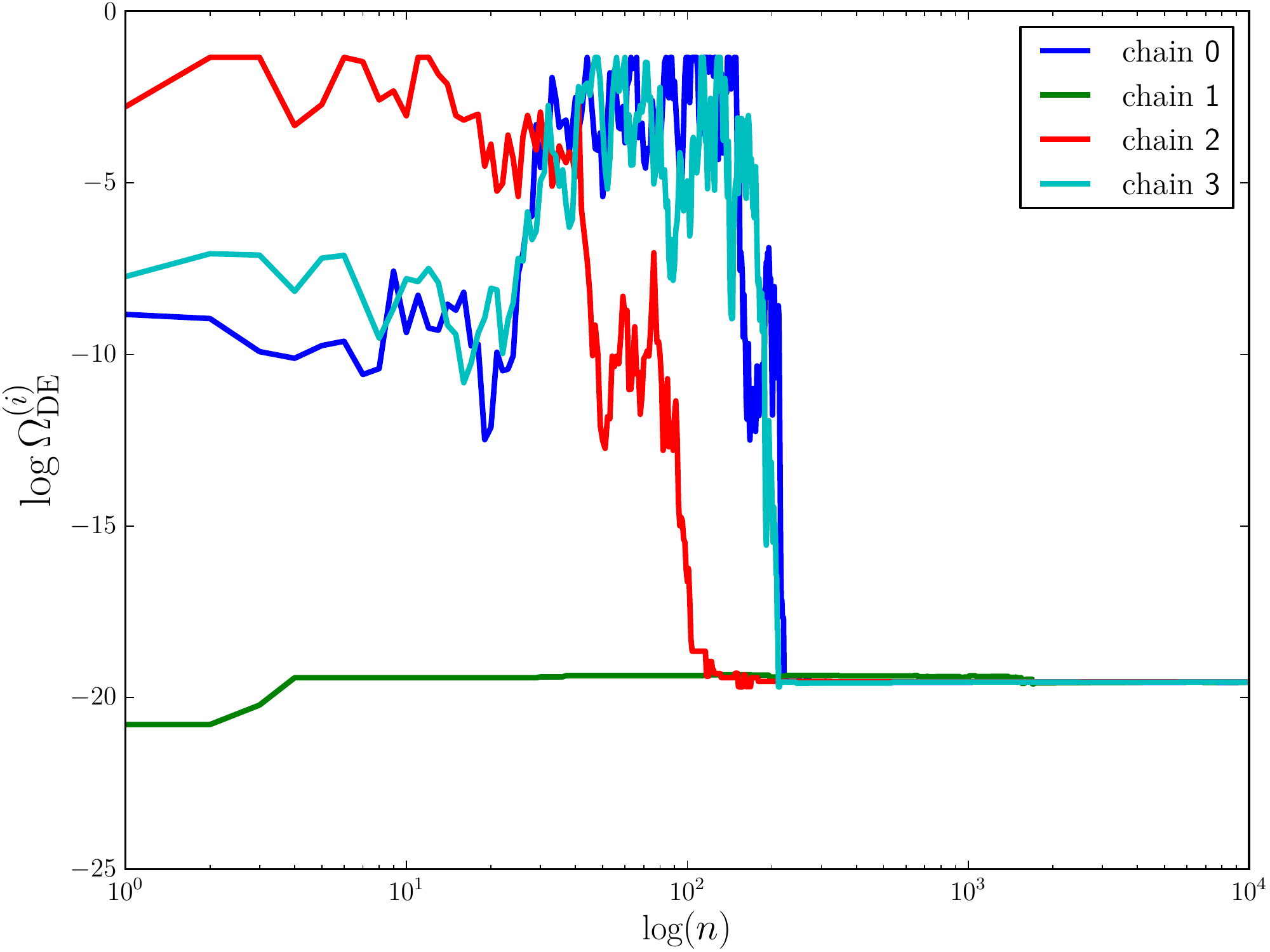}}
\end{minipage}
\hfill
\begin{minipage}[v]{0.5\linewidth}
\center{\includegraphics[width=1.0\linewidth]{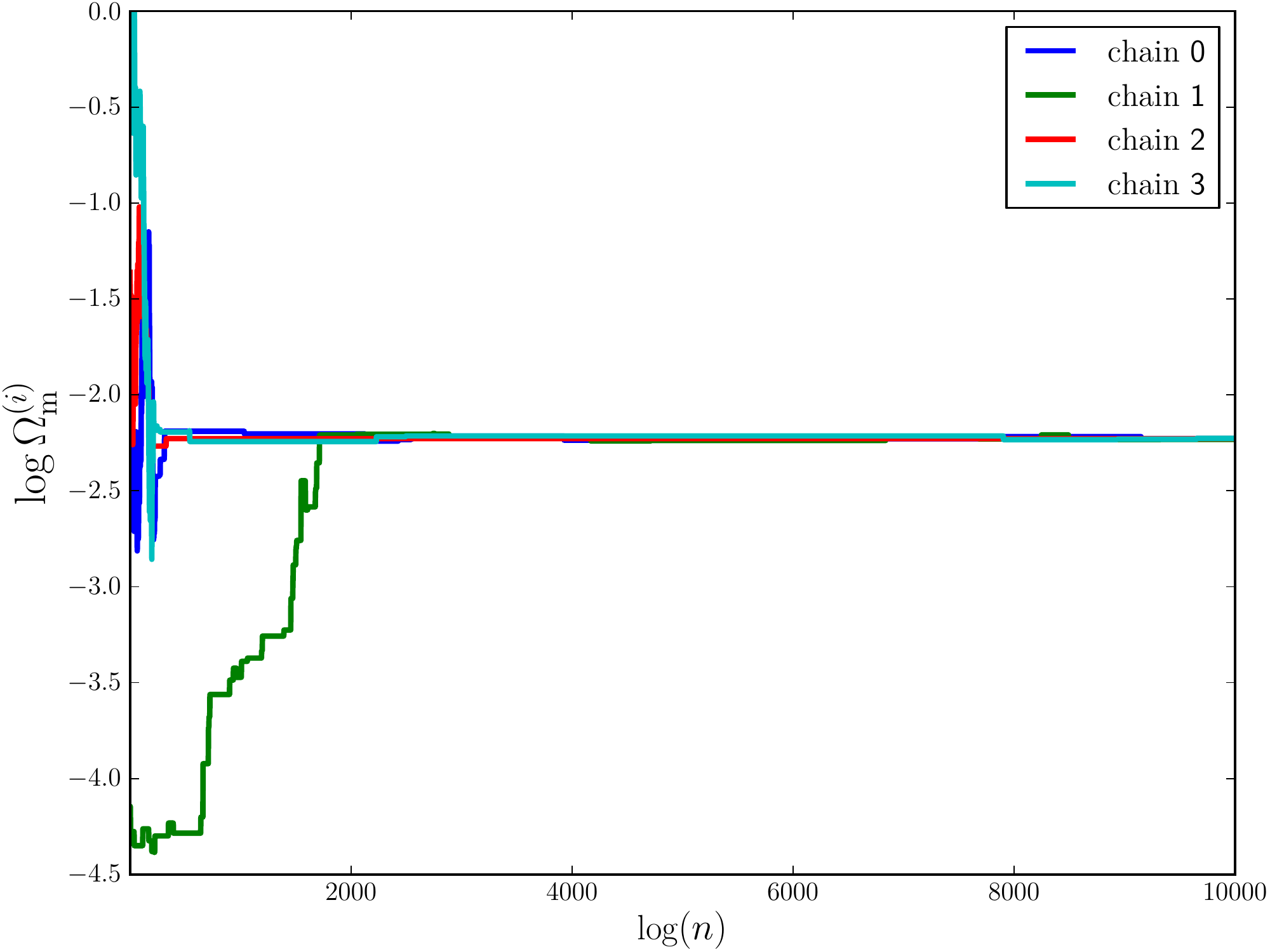}}
\end{minipage}
\caption{MCMC convergence demonstration for
  $\Omega^{(i)}_{DE}$ and $\Omega^{(i)}_{m}$ parameters in the
  $\sigma$CDM model with $\omega^{(i)}_{DE}=0.0$ and $\lambda=0.1$.}
\label{fig:mcmc_parameters_convergence}
\end{figure}

In this paper we concentrated on case when initial value of dark
energy equation of state parameter equals to 0 as in
\cite{Bassett:2007aw, Hwang:2009zj}.

\begin{figure}[h!]
\begin{minipage}[v]{0.5\linewidth}
\center{\includegraphics[width=1.0\linewidth]{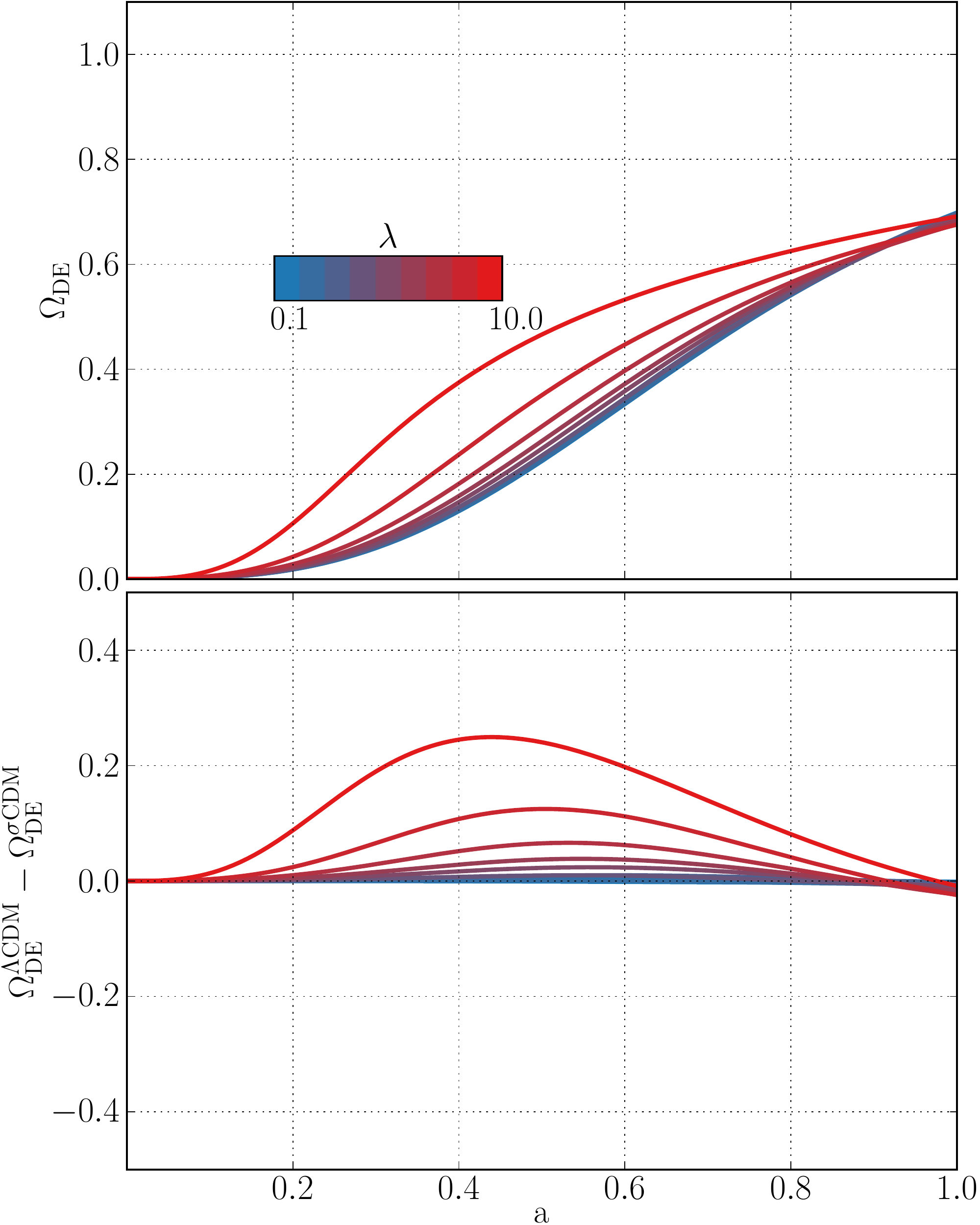}}
\end{minipage}
\hfill
\begin{minipage}[v]{0.5\linewidth}
\center{\includegraphics[width=1.0\linewidth]{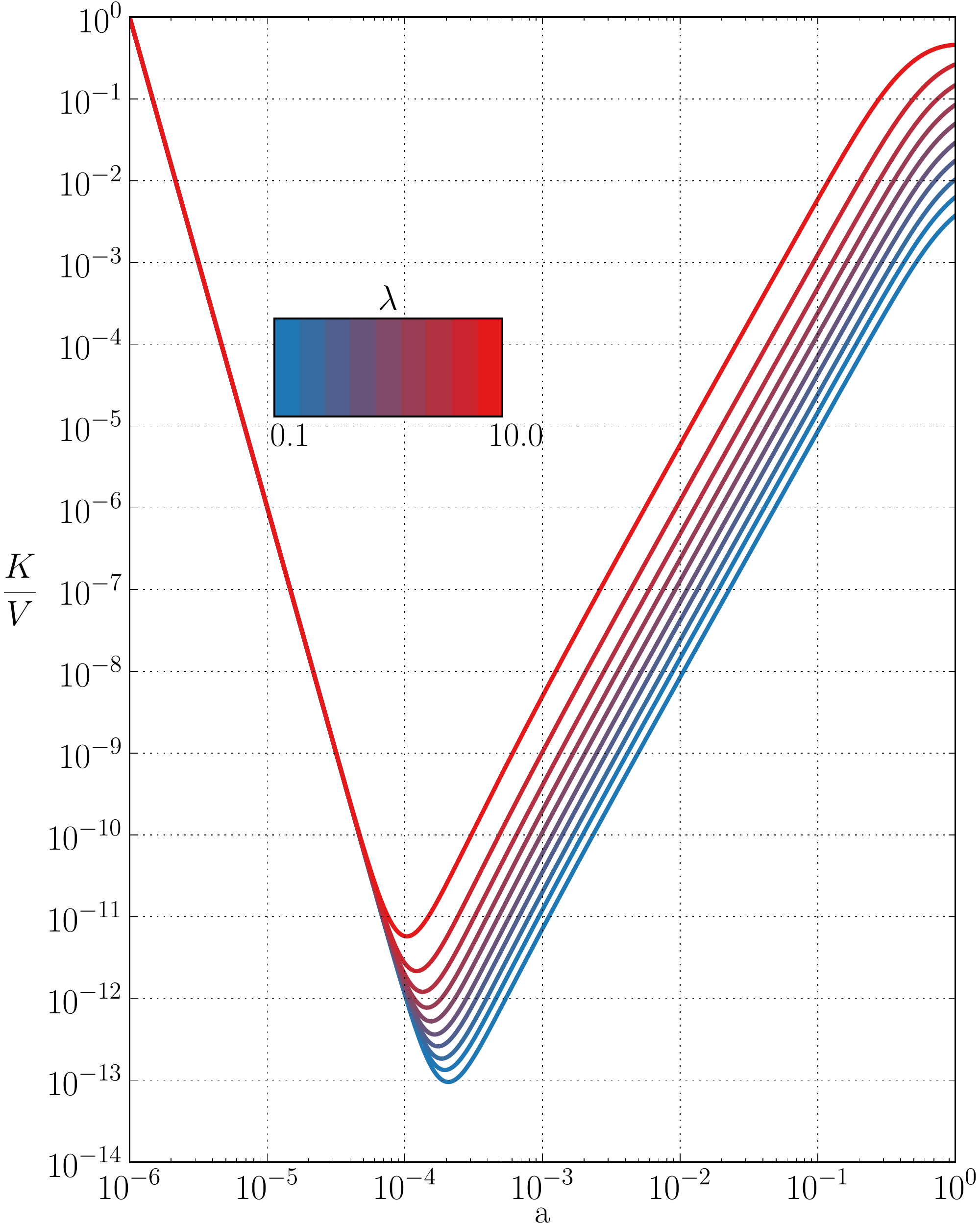}}
\end{minipage}
\caption{The left upper panel is for $\Omega_{DE}$ evolution in $\sigma$CDM
  model and lower left panel is for comparison with
  $\Omega_{DE}$ evolution in the $\Lambda$CDM model. The
  ratio of the kinetic and potential energies of chiral cosmological fields
in $\sigma$CDM model is presented on the right while varying coupling
constant in the potential in $\sigma$CDM}
\label{fig:family_Omega_DE_and_family_K_over_V}
\end{figure}

\begin{figure}[h!]
\begin{minipage}[v]{0.5\linewidth}
\center{\includegraphics[width=1.0\linewidth]{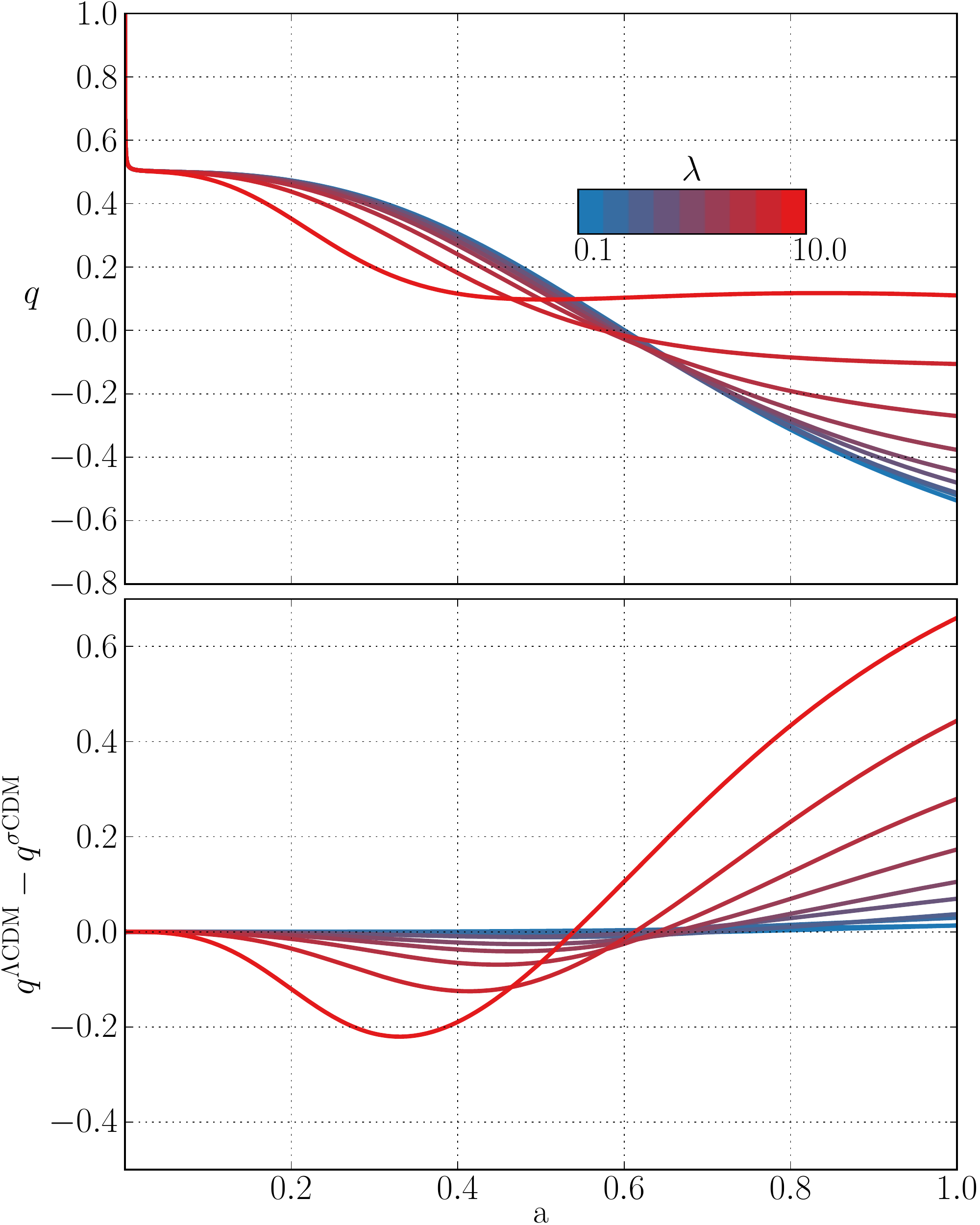}}
\end{minipage}
\hfill
\begin{minipage}[v]{0.5\linewidth}
\center{\includegraphics[width=1.0\linewidth]{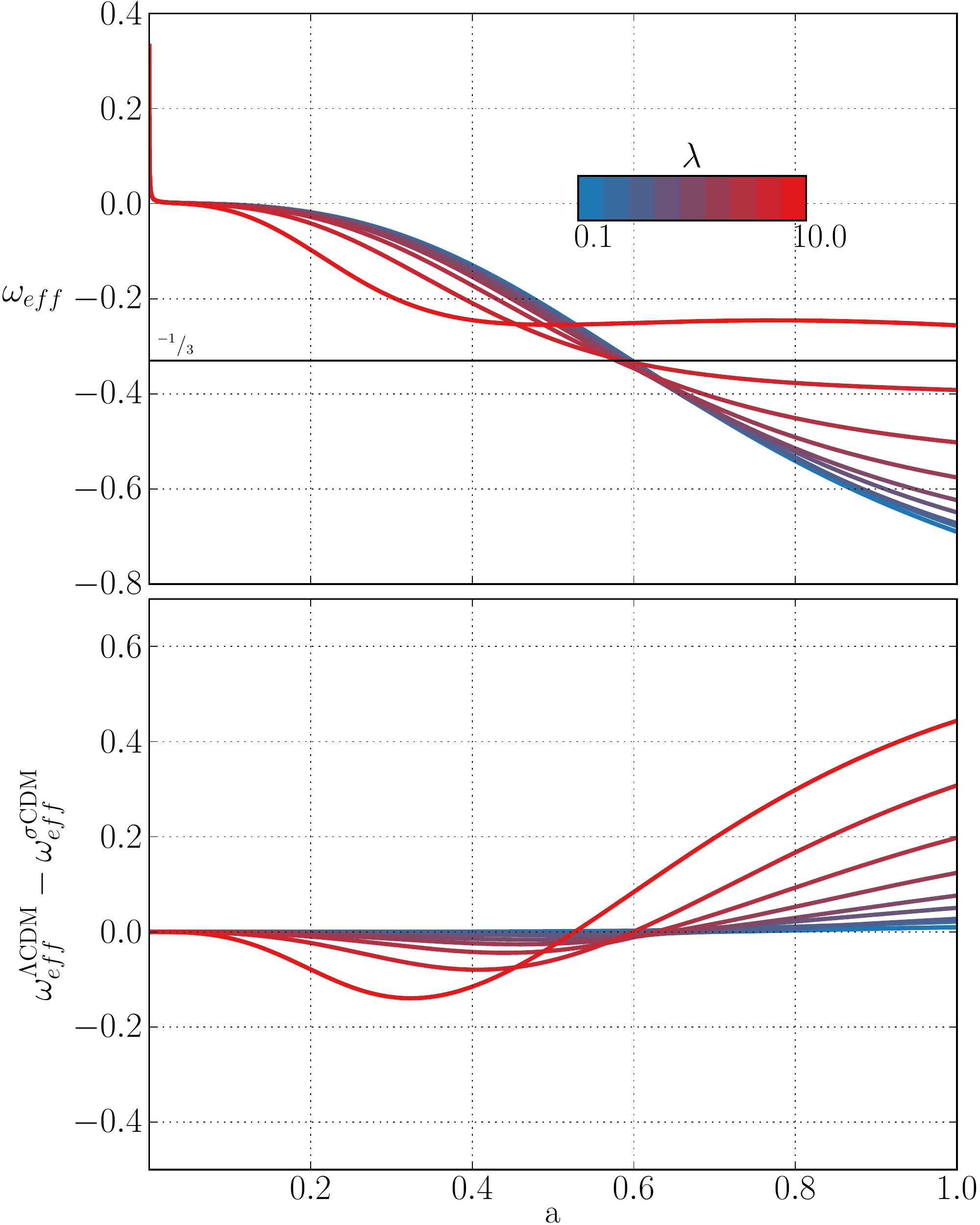}}
\end{minipage}
\caption{The deceleration parameter in the $\sigma$CDM in the left panels in
comparison with evolution $q$ in the $\Lambda$CDM model on the bottom.
The effective equation of state parameter in the $\sigma$CDM is shown on the right in
comparison with evolution $\omega_{eff}$ in the $\Lambda$CDM model in the
bottom right panel.
}
\label{fig:family_q_and_family_omega_eff}
\end{figure}

\begin{figure}[h!]
\center{\includegraphics[width=0.8\textwidth]{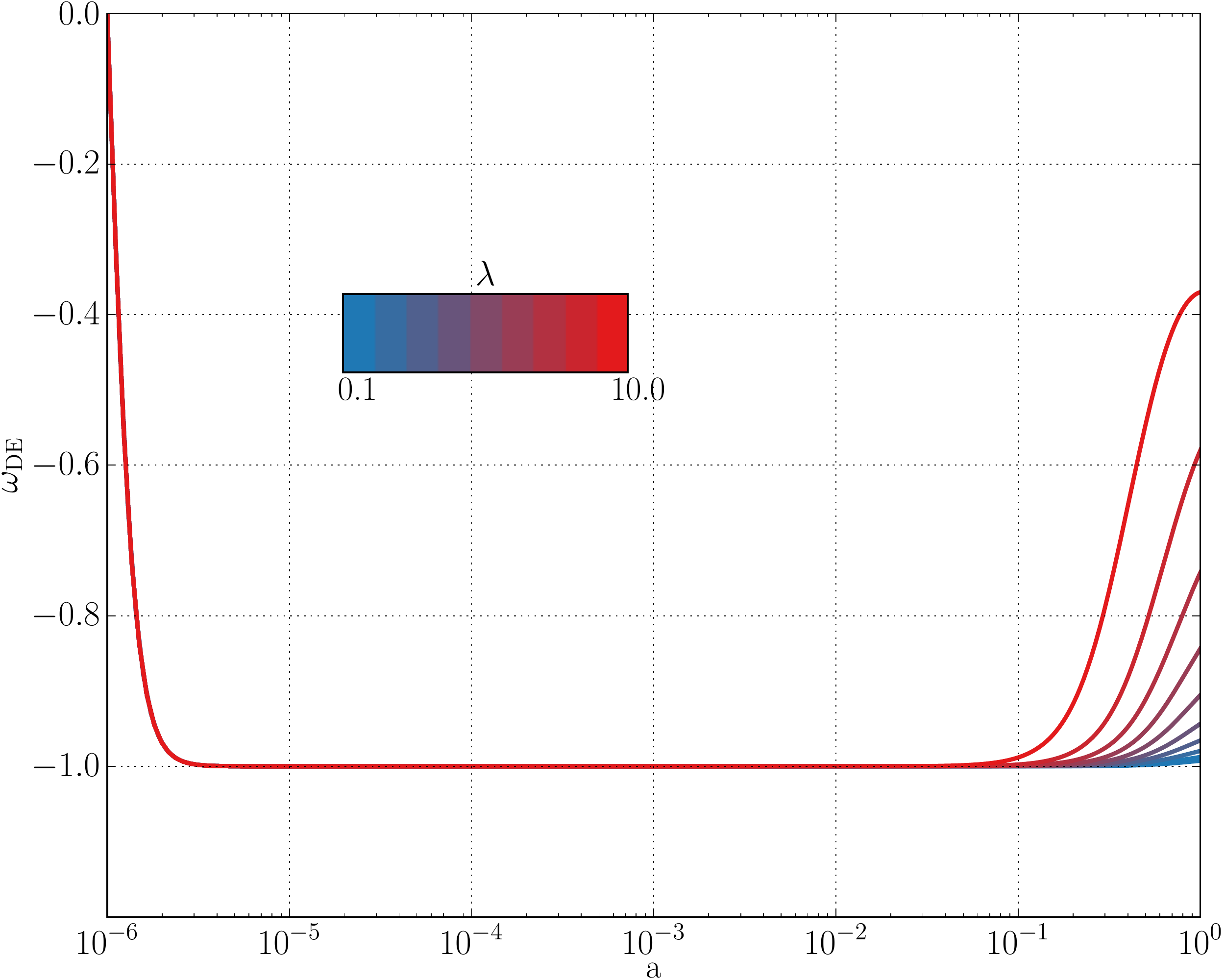}}
\caption{The evolution of the equation of state parameter $\omega_{DE}$ 
in dependence on the potential coupling constant.}
\label{fig:omega_DE_i0.0_w_rad_DE_eq_of_state_param}
\end{figure}

\section{Discussion}
\label{sec:disc}

The model under investigation demonstrates rather interesting behavior
when the potential interaction values is changing. As one can see 
from Fig.\,\ref{fig:family_Omega_DE_and_family_K_over_V}, the 
evolution of dark energy contribution to the 
critical density grows with increasing potential coupling constant $\lambda$. The
smaller is the value of $\lambda$, the less is the deviation of
$\Omega_{DE}$ in the $\sigma$CDM model in comparison with the
$\Lambda$CDM. This fact is in agreement with the general statement that
if we considered limit $\lambda\rightarrow\infty$ than we would have
stiff matter behaviour described by chiral cosmological fields. Therefore 
we notice that for $\lambda=10.0$ we do not get a transition to an accelerated 
expansion.  This is demonstrated in Fig.
\ref{fig:family_q_and_family_omega_eff} for the deceleration parameter $q$
and the effective equation $\omega_{eff}$. In the case
of $\lambda=10.0$, the solutions for $q$ and $\omega_{eff}$ do not cross 0
and $-{}^{1}/{}_3$ values, respectirely. We do not
observe significant contributions of dark energy at early
epochs which could suppress radiation component at
previous epochs of the cosmic evolution.
 
It is interesting to study relation between kinetic and potential
interactions in subsequent stages of cosmic evolution, see
Fig.\,\ref{fig:family_Omega_DE_and_family_K_over_V}.  An initial value
of $\omega^{(i)}_{DE}=0$ leads to $K=V$ at this initial time.  We have
also found that the value $h_{22}$ is equal to $1$ almost all time and
deviates somewhat in recent times, so we have a pure canonical
2--component model with very small influence from chiral metric
coefficient.
 
We should add a few words about the behavior of dark energy equation
of state $\omega_{DE}$ described by chiral cosmological fields EoS
parameter. We found that it drops down almost immediately to value of
$-1$ for all models under investigation and only at recent times there
is some distinction between models with different values of $\lambda$
as seen in Fig.  \ref{fig:omega_DE_i0.0_w_rad_DE_eq_of_state_param}.
As we mentioned before the $\sigma$CDM model with $\lambda \ge 10$
should be excluded as dark energy model, and the value $\lambda=5.99$
is acceptable from our criteria for viable cosmological model.
Distinguishing between the remaining 9 parameters values requires additional
consideration and additional observational data.
 
\section{Conclusion and future work}
\label{sec:conclusion}

In this article we investigated the $\sigma$CDM model minimally
coupled with radiation. The MCMC procedure allowed us to effectively
evaluate the parameters space. We have found that the kinetic
interaction $h_{22}$ has to be equal to $1$ nearly all time except at
the period close to the present one. To avoid this, the simple form
chosen in this article should be replaced by a more complicated one.
The universe described by model under consideration shows a behavior
very similar to hat of the $\Lambda$CDM model if potential interaction
coupling constant $\lambda$ is less than $1$, and it deviates strongly
when this parameter increases. It was found that models with large
$\lambda$ does not describe an accelerated Universe expansion. This
restriction may be avoided by taking into account variation of others
model parameters.  These issues we plan to cover in a future
publication.

\acknowledgments

SVC and RRA are supported by State Order of the Ministry of
Education and Science of the Russian Federation in accordance with
Project No.2014/391. SVC and RRA also would like to thank
Leibniz Institut f\"ur Astrophysik Potsdam for warm hospitality when part
of this article was prepared.

\providecommand{\href}[2]{#2}\begingroup\raggedright\endgroup


\end{document}